\documentclass{pasj00}

\begin{document}
\SetRunningHead{M. Honma et al.}{W 49N H$_2$O Maser Outburst in 2003 October}
\Received{2004/2/25}
\Accepted{2004/4/22}

\title{VERA Observation of the W 49N H$_2$O Maser Outburst in 2003 October}


\author{
Mareki \textsc{Honma},\altaffilmark{1,2,3}
Yoon Kyung \textsc{Choi},\altaffilmark{1,4}
Takeshi \textsc{Bushimata},\altaffilmark{1,5}
Takahiro \textsc{Fujii},\altaffilmark{1,6}
Tomoya \textsc{Hirota},\altaffilmark{1,2}\\
Koji \textsc{Horiai},\altaffilmark{1,7}
Hiroshi \textsc{Imai},\altaffilmark{6}
Noritomo \textsc{Inomata},\altaffilmark{6}
Jose \textsc{Ishitsuka},\altaffilmark{1}
Kenzaburo \textsc{Iwadate},\altaffilmark{1,7}\\
Takaaki \textsc{Jike},\altaffilmark{1}
Osamu \textsc{Kameya},\altaffilmark{1,3,7}
Ryuichi \textsc{Kamohara},\altaffilmark{1,6}
Yukitoshi \textsc{Kan-ya},\altaffilmark{1}\\
Noriyuki \textsc{Kawaguchi},\altaffilmark{1,2,3}
Hideyuki \textsc{Kobayashi},\altaffilmark{1,5}
Seisuke \textsc{Kuji},\altaffilmark{1,2}
Tomoharu \textsc{Kurayama},\altaffilmark{1,4}\\
Seiji \textsc{Manabe},\altaffilmark{1,2,3}
Takeshi \textsc{Miyaji},\altaffilmark{1,5}
Akiharu \textsc{Nakagawa},\altaffilmark{6}
Kouichirou \textsc{Nakashima},\altaffilmark{6}\\
Riiko \textsc{Nagayoshi},\altaffilmark{6}
Toshihiro \textsc{Omodaka},\altaffilmark{6}
Tomoaki \textsc{Oyama},\altaffilmark{1,4}
Maria \textsc{Rioja},\altaffilmark{1}
Satoshi \textsc{Sakai},\altaffilmark{1,2}\\
Sei-ichiro \textsc{Sakakibara},\altaffilmark{6}
Katsuhisa \textsc{Sato},\altaffilmark{1,7}
Tetsuo \textsc{Sasao},\altaffilmark{1,8}
Katsunori M. \textsc{Shibata},\altaffilmark{1,2}\\
Rie \textsc{Shimizu},\altaffilmark{6}
Kasumi \textsc{Sora},\altaffilmark{6}
Hiroshi \textsc{Suda},\altaffilmark{1,4}
Yoshiaki \textsc{Tamura},\altaffilmark{1,2,3} and 
Kazuyoshi \textsc{Yamashita}\altaffilmark{6}
}

\altaffiltext{1}{VERA Project Office, NAOJ, Mitaka, Tokyo 181-8588}
\altaffiltext{2}{Earth Rotation Division, NAOJ, Mizusawa, Iwate 023-0861}
\altaffiltext{3}{Graduate University for Advanced Studies, Mitaka, Tokyo 181-8588}
\altaffiltext{4}{Department of Astronomy, University of Tokyo, Bunkyo, Tokyo 113-8654}
\altaffiltext{5}{Radio Astronomy Division, NAOJ, Mitaka, Tokyo 181-8588}
\altaffiltext{6}{Faculty of Science, Kagoshima University, Korimoto, Kagoshima, Kagoshima 890-0065}
\altaffiltext{7}{Mizusawa Astro-Geodynamics Observatory, NAOJ, Mizusawa, Iwate 023-0861}
\altaffiltext{8}{Ajou University, Suwon 442-749, Republic of Korea}

\email{honmamr@cc.nao.ac.jp}

\maketitle

\begin{abstract}

We report on a strong outburst of the W 49N H$_2$O maser observed with VERA.
Single-dish monitoring with VERA 20 m telescopes detected a strong outburst of the maser feature at $V_{\rm LSR}$ = $-30.7$ km s$^{-1}$ in 2003 October.
The outburst had a duration of $\sim$ 100 days and a peak intensity of 7.9$\times 10^4$ Jy, being one of the strongest outbursts in W 49N observed so far.
VLBI observations with the VERA array were also carried out near to the maximum phase of the outburst, and the outburst spot was identified in the VLBI map.
While the map was in good agreement with previous studies, showing three major concentrations of maser spots, we found a newly formed arc-like structure in the central maser concentration, which may be a shock front powered by a forming star or a star cluster.
The outburst spot was found to be located on the arc-like structure, indicating a possible connection of the present outburst to a shock phenomenon.

\end{abstract}

\section{Introduction}

Water maser emission from star-forming regions is known to show a strong time variation, and the most notable variations are referred to as {`}outbursts{'}, in which a maser feature with a narrow line width shows dramatic brightening by more than an order of magnitude with a duration of a few tens of days to a few months.
Although the outburst mechanism is still puzzling, observations of outbursts provide us with a unique opportunity to study maser physics as well as the physical condition of star-forming regions that harbor maser clouds.
Most intensive studies of maser outburst have been carried out in the two major H$_2$O maser sources, namely Orion KL, one of the nearest massive-star forming regions from the Sun, and W 49N, the brightest H$_2$O maser source in the Milky Way Galaxy.
As for the Orion KL maser, intensive studies were conducted for strong outbursts that occurred in 1979 and 1997 (Matveenko et al. 1980; Omodaka et al. 1999).
VLBI observations of the outburst in 1997 (e.g., Matveyenko et al. 1998; Kobayashi et al. 2000) revealed a structure evolution during the peak phase, indicating a maser--maser interaction origin, and the polarization measurement of the outburst suggested that the phenomenon is also related to a strong magnetic field (Horiuchi, Kameya 2000).
On the other hand, outbursts of the W49 N maser have been studied mainly based on single-dish monitoring of the flux variation (Boboltz et al. 1998; Liljestr\"om, Gwinn 2000; Zhou et al 2002).
These monitoring investigations showed that W 49N also exhibits H$_2$O maser outbursts at various radial velocities, with the most prominent one occurring in 1983 with a peak flux of $8.9\times 10^4$ Jy (Liljestr\"om, Gwinn 2000).
Recent single-dish studies of W 49N revealed small, but systematic, changes in the radial velocity and the velocity width during outbursts, and sophisticated maser--maser interaction models were proposed to explain the observed properties (Boboltz et al. 1998; Zhou et al. 2002).

In 2003 October, another strong outburst occurred in W 49N, and we detected the outburst during the course of test observations of VERA.
The outburst had a peak intensity of 7.9$\times 10^4$ Jy, being one of the strongest outbursts observed in W 49N.
We also carried out VLBI observation of W 49N to identify the outburst spot, since high-quality VLBI maps of W 49N were only available for observations in the early 1980's (e.g., Walker et al. 1982; Gwinn et al. 1992).
In this letter we present the single-dish properties of the outburst as well as a VLBI map obtained during the outburst in 2003 October.
We compare the characteristics of the previous and present outbursts, and discuss the possible mechanism of the present outbursts.

\section{Observations and Reductions}

\subsection{Single-Dish Observations}

As a series of test observations of VERA, monitoring of the W 49N H$_2$O maser was performed with an interval of 2 to 3 weeks, most of which are snap-shot VLBI observations for 10 to 20 minutes to check the fringe detection and system stability.
Among them, we used the auto-correlation data obtained at Mizusawa or Iriki station (a station with a better condition was used).
Auto-correlation spectra were processed with Mitaka FX correlator with a 16 MHz bandwidth and 512 spectral channels, yielding frequency and velocity resolutions of 37.5 kHz and 0.42 km s$^{-1}$, respectively.
Spectral analyses were performed in the usual manner: an amplitude calibration was performed based on position switching with the system noise temperature measured by the R-Sky method, and observed frequency was converted to the velocity with respect to the Local Standard of Rest (LSR) using a rest frequency of 22.235080 GHz.
Among the data obtained in 2003, here we present the data collected on the days of year 234 (August 22), 258 (September 15), 281 (October 8), 294 (October 21), and 325 (November 21), which cover almost the whole duration of the outburst.

\subsection{VLBI Observation}

A VLBI observation of W 49N was performed on day 294 with 3 stations of VERA (Mizusawa, Iriki, and Ishigaki-jima) for 8 hours. 
As a part of VERA's phase-referencing performance test (for details see Honma et al. 2003), the observation was made in the dual-beam mode, and another H$_2$O maser source, OH43.8$-$0.1 (0$^\circ$.65 separation from W 49N), was observed simultaneously.
A bright continuum source, TXS 1923+210, was also observed every 2 hours as a clock and bandpass calibrator.
The VSOP-terminal system was used as a digital back-end, and digitized data were recorded onto magnetic tapes at a data rate of 128 Mbps with 2-bit quantization.
Among the total bandwidth of 32 MHz, one 16 MHz channel was assigned to W 49N, covering the LSR velocity range between $-113.9$ and $101.4$ km s$^{-1}$.
Correlation processing was performed with the Mitaka FX correlator with a spectral resolution of 512 points per channel, yielding the frequency and velocity resolutions of 37.5 kHz and 0.42 km s$^{-1}$, which are the same as those of auto-correlation data.
In the data analysis, visibilities of all velocity channels were phase-referenced to the reference maser spot at $V_{\rm LSR}$ of 9.1 km s$^{-1}$, which is one of the brightest spots, and shows no sign of structure according to the closure phase.
Phase-referenced visibilities were Fourier transformed to synthesize images, and the positions of the brightness peaks were determined with respect to the reference spot.
The synthesized beam has a FWHM beam size of 1.2$\times$2.3 mas with a PA of 45$^\circ$.

\begin{figure}
\begin{center}
	\includegraphics[clip,width=11.7cm,angle=-90]{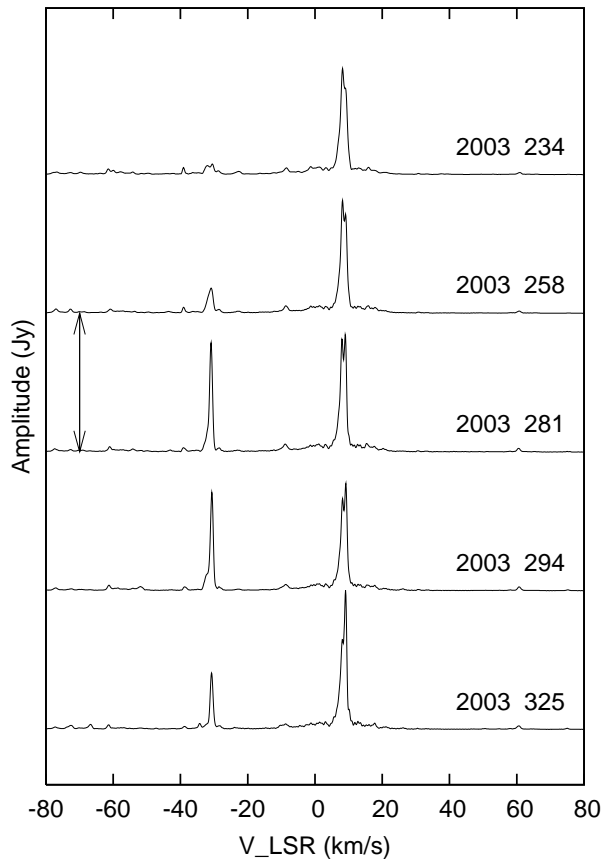}
\end{center}
\caption{Single-dish spectral evolution of W 49N H$_2$O maser from August to November in 2003. 
The vertical arrow shows the scale of $10^5$ Jy, and each spectrum is vertically shifted by the same amount.
While the maser feature at 8--9 km s$^{-1}$ is stable during 5 epochs, the feature at $V_{\rm LSR}$=$-30.7$ km s$^{-1}$ shows a strong outburst with a peak intensity of 7.9$\times 10^4$ Jy.}
\end{figure}

\section{Results}
\subsection{Single-Dish Spectral Evolution}

Figure 1 shows the spectral evolution of W 49N observed with VERA.
In the five epochs shown figure 1, the brightest feature at $V_{\rm LSR}$=8--9 km s$^{-1}$ was quite stable, being as bright as $10^5$ Jy.
In contrast, the maser feature at $V_{\rm LSR}$=$-30.7$ km s$^{-1}$ showed a dramatic variation of the flux density.
On day 234, the outburst was not evident, with a flux density of less than 8$\times 10^3$ Jy.
On day 258 the outburst feature showed a sign of flaring.
The outburst reached the maximum value of 7.9$\times 10^4$ Jy on day 281, and on day 294 the outburst feature was still prominent.
One month later (on day 325) the outburst seemed to be declining, with its flux density dropping to about half of the maximum, being 4.1$\times 10^4$ Jy.

During the bright phase (on day 281, 294, and 325), the outburst feature showed no variation of the line width, being 0.96$\pm$0.04 km s$^{-1}$ (Full Width at Half Maximum).
This is a typical value when compared with previous flares (e.g., Liljestr\"om, Gwinn 2000).
The peak velocity was $-30.7$ km s$^{-1}$ on days 294 and 325, while $-30.9$ km s$^{-1}$ on day 281.
The peak velocity shifts were also reported in previous outbursts (Liljestr\"om \& Gwinn 2000; Boboltz et al. 1998; Zhou et al. 2002), but our data samplings were not high enough to see if any systematic velocity drift or velocity jump had occurred between day 281s and 294.

Liljestr\"om and Gwinn (2000) reported an outburst of $V_{\rm LSR}=-30.4$ km s$^{-1}$ feature, which is likely to be the same feature as that of the present outburst.
However, the previous outburst had a maximum intensity of 1.2$\times 10^4$ Jy, which is 6.5-times smaller than the present one.
The peak flux of the present outburst, 7.9$\times 10^4$ Jy, is comparable to that of the strongest one out of 146 outbursts in Liljestr\"om and Gwinn (2000), which occurred at $V_{\rm LSR}=-0.9$ km s$^{-1}$ with a peak flux of 8.9$\times 10^4$ Jy.
Liljestr\"om and Gwinn (2000) also reported that typical outbursts of maser features around $V_{\rm LSR}=-30$ km s$^{-1}$ tended to have a peak flux of $\sim 3\times 10^3$ Jy.
The peak flux of the present outburst, 7.9$\times 10^4$ Jy, is rather exceptional when compared with the previous studies.
For instance, the outburst in the Orion KL H$_2$O maser in 1997 had a peak intensity of 5$\times 10^6$ Jy with a FWHM line width of 0.40 km s$^{-1}$ (Omodaka et al. 1999; Kobayashi et al. 2000).
Adopting the distances to W 49N of 11.4 kpc (Gwinn et al. 1992) and to Orion KL of 480 pc (Genzel et al. 1981), the isotropic luminosity of the present outburst was 22-times larger than that of the Orion KL outburst in 1997.
Thus, the present outburst is not only one of the brightest outbursts in W 49N, but also the most prominent one among the major maser sources in the Milky Way Galaxy.

\subsection{VLBI Map}

Figure 2 shows the maser spot distribution in W 49N: the left panel shows the global distribution of the maser spots on the scale of a few arcsec, and the right panel shows an expanded map of the maser concentration around ($X$, $Y$)=(0 mas, 0 mas).
As described in the previous section, the map origin is identical to the location of the reference spot at $V_{\rm LSR}$=9.1 km s$^{-1}$.
In total, 341 maser spots have been identified, and 329 spots are shown in figure 2 (other 12 spots are out of the plot range of figure 2).
The spot distribution in figure 2 is in good agreement with that of previous studies (e.g., Walker et al. 1982; Gwinn et al. 1992), showing three major concentrations of maser spots (the central one at $X\sim 0^{''}$ with north--south extension of 600 mas, the eastern one at $X\sim 1^{''}$, and the western one at $X\sim -0.5^{''}$).
The radial velocity structure, which shows a redshifted eastern concentration and a blueshifted western concentration, is also in good agreement with previous studies.

While the global distributions of maser spots are consistent with the maps obtained about 20 years ago, there is a remarkable difference between the previous and present VLBI maps.
As can be seen in the right panel of figure 2, the central concentration shows an arc-like structure in the southern part, with $X$ from $-30$ to 80 mas and $Y$ from $-280$ to 30 mas.
In the previous maps, the central maser concentration showed a rather straight structure with a north--south extension (Walker et al. 1982; Gwinn et al. 1992) and this kind of arc-like structure was not seen, indicating that the arc-like structure was newly formed during the past 20 years.
Interestingly, our VLBI map shows that the present outburst occurred on this arc-like structure.
The relative position of the outburst spot at $V_{\rm LSR}=-30.7$ km s$^{-1}$ was measured as ($X$, $Y$)=(69.67 mas, $-68.55$ mas).
The position of the outburst spot is indicated with an arrow in the right panel of figure 2.
The outburst location on the arc-like structure indicates a possible link between the maser outburst phenomenon and shock in the star-forming regions (see discussion).

Although a good spot image is not available based on our observations (only 3 baselines), we can obtain some constraint on the maser feature structure.
According to the closure phase, there is no sign of a structure larger than the synthesized beam.
Also, the positions of spots in the neighboring channels ($V_{\rm LSR}=-29.6$ km s$^{-1}$ to $-31.8$ km s$^{-1}$) coincide with the position of the outburst peak at $V_{\rm LSR}=-30.7$ km s$^{-1}$ with an accuracy of 0.1 mas.
Thus, the outburst spot feature is likely to have a size of less than 1.1 AU (0.1 mas in angular scale).
However, we note that the apparent lack of structure in the outburst spot does not necessarily rule out the maser--maser interaction mechanism for the outburst origin.
For example, Kobayashi et al. (2000) found a double-peaked structure of the Orion KL outburst spot on a physical scale of about 1 AU, which is fairly close to the upper limit on the spot size for the present W 49N outburst.

\begin{figure}
\begin{center}
	\includegraphics[clip,width=13.2cm,angle=-90]{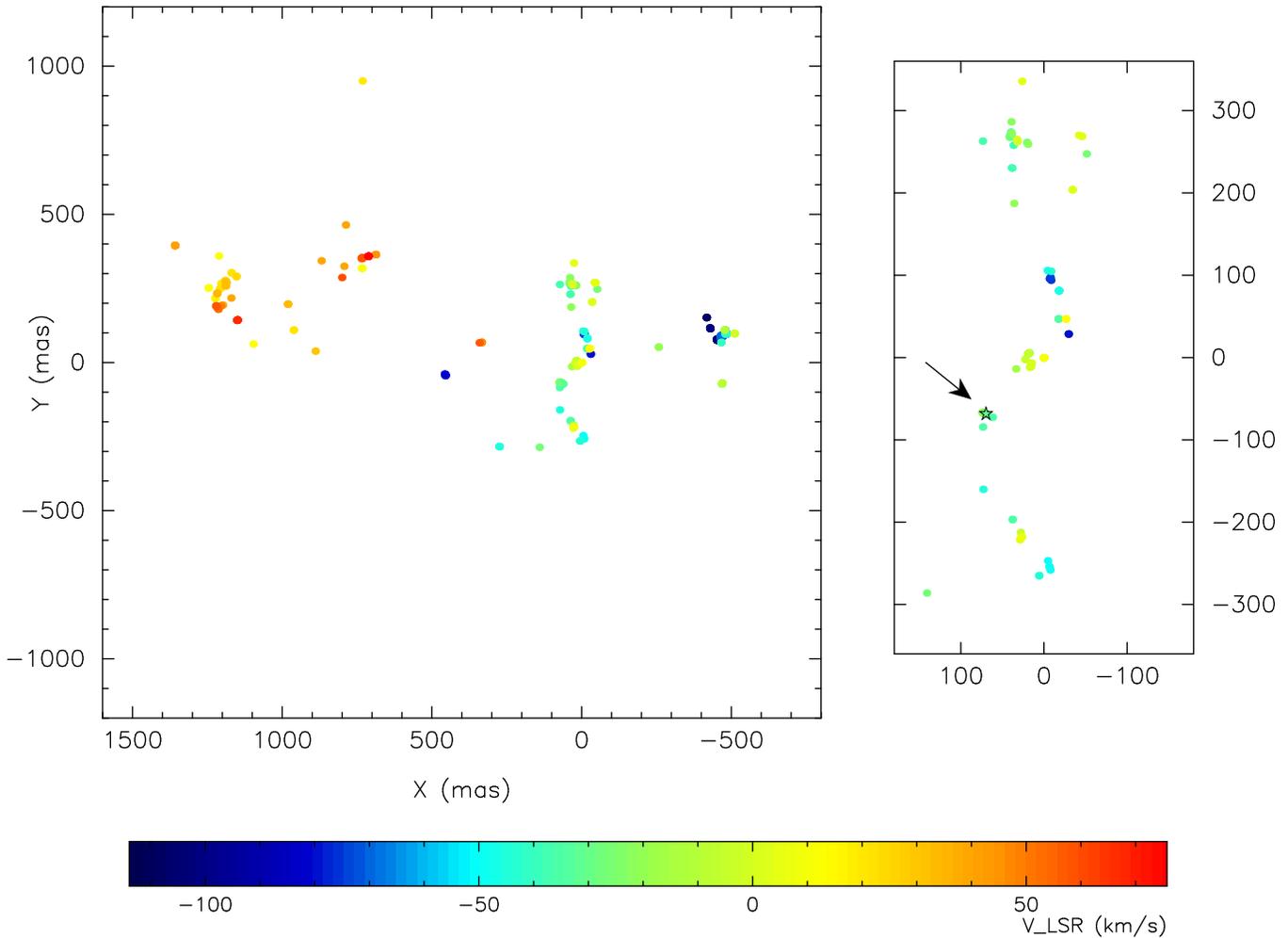}
\end{center}
\caption{Maser spot distribution in W 49N. 
Left panel shows the spot distribution on the scale of a few arcsec.
Right panel is the extended map of the central maser concentration.
The outburst spot at $V_{\rm LSR}=-30.7$ km s$^{-1}$ is indicated with an arrow and a star.
Spot color indicates the radial velocity of each spot (see color index at the bottom).}
\end{figure}

\section{Discussion}
As can be seen in the previous section, our VLBI map revealed that the outburst spot is located on the arc-like structure.
Although it is not clear how the arc-like structure was formed, the superposition of Gwinn et al.'s map onto our map implies that a part of the north--south maser extension seen in Gwinn et al.'s map was pushed toward the east, and the arc-like structure currently seen was formed.
If this interpretation is correct, the arc-like structure is likely to trace the shock front caused by the gas-flow motion (like a shock shell in Cep A, Torrelles et al. 2001).
In that case, the outburst location suggests that the present outburst is tightly related to shock phenomenon, as was studied previously (Liljestr\"om, Gwinn 2000).
Since our observation shows no structure in outburst spots, the outburst may be simply caused by a change in the physical properties of the shock regions.

As we have already mentioned, we cannot rule out the maser--maser interaction scheme, whose size is smaller than our angular resolution.
Nevertheless, if the shock-front interpretation is correct, at least one can rule out a maser--maser interaction in a circumstellar disk (Matveenko et al. 1988), since the present outburst is located in a shock front rather that in a circumstellar disk.
The other interesting feature of the shock-front interpretation is that this indicates the existence of an activity center other than the common kinematic center obtained by Gwinn et al.(1992).
Although the common center position by Gwinn et al.(1992) is somewhat model-dependent, their results favor a common center in the central maser concentration, or east of the central maser concentration.
On the other hand, our shock-front interpretation requires another outflow origin west of the central maser concentration.
In order to verify the shock-front interpretation, measuring the proper motion is one of the most powerful approaches, which will be our next step for a deeper understanding of maser phenomena in W 49N.

\vspace{1pc}\par
One of the authors (MH) acknowledges financial support by Inamori Foundation.
Part of the data reduction was performed at the Astronomical Data Analysis Center of the NAOJ, which is an inter-university research institute of astronomy operated by Ministry of Education, Culture, Sports, Science and Technology.\\
\vspace*{13cm}



\begin{thebibliography}{}

\bibitem[Boboltz et al.(1998)]{}
Boboltz, D. A., et al. 1998, ApJ, 509, 256

\bibitem[Genzel et al.(1981)]{}
Genzel, R., Reid, M. J., Moran, J. M., \& Downes D. 1981, ApJ, 244, 884

\bibitem[Gwinn et al.(1992)]{}
Gwinn, C. R., Moran, J. M., \& Reid, M. J. 1992, ApJ, 393, 149

\bibitem[Honma et al.(2003)]{}
Honma, M., et al. 2003, PASJ, 55, L57

\bibitem[Horiuchi, Kameya(2000)]{}
Horiuchi, S., \& Kameya, O. 2000, PASJ, 52, 545

\bibitem[Kobayashi et al.(2000)]{}
Kobayashi, H., et al. 2000 in Astrophysical Phenomena Revealed by Space VLBI, ed H. Hirabayashi, P. G. Edwards, \& D. W. Murphy (Sagamihara: ISAS), 109

\bibitem[Matveenko et al.(1980)]{}
Matveenko, L. I., Kogan, L. R., \& Kostenko, V. I. 1980, Sov. Astron. Lett., 6, 279

\bibitem[Matveenko et al.(1988)]{}
Matveyenko, L. I., Graham, D. A., \&Diamond, P. J.  1988, Sov. Astron. Lett., 14, 468

\bibitem[Matveenko et al.(1998)]{}
Matveenko, L. I., Diamond, P. J., \& Graham, D. A.  1998, Astron. Lett., 24, 623

\bibitem[Liljestr\"om, Gwinn (2000)]{}
Liljestr\"om T., \& Gwinn, C. R. 2000, ApJ, 534, 781

\bibitem[Omodaka et al.(1999)]{}
Omodaka, T., et al. 1999, PASJ, 51, 333

\bibitem[Torrelles et al.(2001)]{}
Torrelles, J. M., et al. 2001, Nature, 411, 277

\bibitem[Walker et al.(1982)]{}
Walker, R. C., Matsakis, D. N., \& Garcia-Barreto, J. A. 1982, ApJ, 255, 128

\bibitem[Zhou et al.(2002)]{}
Zhou, J. J., Zheng, X. W., \& Chen, Y. X. 2002, A\&A, 390, 281

\end{thebibliography}
\end{document}